\documentclass[a4paper,man,natbib]{apa6}

\usepackage{booktabs} % For formal tables
\usepackage{array}
\usepackage{apacite}
\usepackage{longtable}
\usepackage{setspace}
\usepackage[ruled]{algorithm2e} % For algorithms
\usepackage{tabularx}
\usepackage{color}

\newcommand{\cut}[1]{{}}

\title {Crafting Moral Infrastructures: How Nonprofits Use Facebook to Survive}
\shorttitle{How Nonprofits Use Facebook to Survive}
\author{Libby Hemphill,\textsuperscript{*} \ A.J. Million,\textsuperscript{*} \ Ingrid Erickson\textsuperscript{**}}
\affiliation{University of Michigan School of Information,\textsuperscript{*} \ Syracuse University School of Information Studies\textsuperscript{**}}

\begin{document}
\maketitle
\begin{abstract}

We present findings from interviews with 23 individuals affiliated with non-profit organizations (NPOs) to understand how they deploy information and communication technologies (ICTs) in civic engagement efforts. Existing research about NPO ICT use is largely critical, but we did not find evidence that NPOs fail to use tools effectively. Rather, we detail how various ICT use on the part of NPOs intersects with unique affordance perceptions and adoption causes. Overall, we find that existing theories about technology choice (e.g., task-technology fit, uses and gratifications) do not explain the assemblages NPOs describe. We argue that NPOs fashion infrastructures in accordance with their moral economy frameworks rather than selecting tools based on utility. Together, the rhetorics of infrastructure and moral economies capture the motivations and constraints our participants expressed and challenge how prevailing theories of ICT usage describe the non-profit landscape. 
\end{abstract}

\section{Introduction}\label{introduction}

As individuals and their communities blur the lines between online and
offline life, nonprofit organizations (NPOs) increasingly choose to extend
their civic engagement efforts via information and communication
technologies (ICTs) like social media
\citep{Mansfield2018-by}. They do this because
civically engaged communities have been shown to experience lower rates
of crime, poverty, and unemployment, and have better health and
education than their less engaged counterparts
\citep{Beggs1996-nk,Norris2001-ep,Xenos2007-aq}. What
does it mean for a community to be ``engaged'' and how does this
engagement happen? Generally, activities that pertain to the life of a
community ``count'' as civic engagement. Researchers often provide
examples such as poker clubs and bowling leagues when writing about
civic engagement, but explicitly political activities such as voting and
working with labor unions also appear in the literature
\citep{Putnam2001-sg,Skocpol2004-pa,Norris2001-ep}.
Given the broad array of civic engagement activities in which NPOs engage, it is not surprising that the
decisions NPOs make about which technologies to employ reflects
a set of equally broad goals.

Researchers have begun studying their technology choices and related
practices as nonprofits use ICTs to engage with their local communities \citep{Li2018-gi,Bopp2017-nh,Kim2014-qg,Voida2011-tj,Hou2015-gq}.
This work showcases the breadth of NPOs civic engagement activity, ranging from the use of Twitter to encourage social action \citep{Li2018-gi} to building a strong organization by creating bespoke technology assembalges \citep{Voida2011-tj}. While rich in empirical detail, much of this work is critical, often constructively so, when it comes to NPOs' online engagement efforts. For example, social media use by NPOs is shown to produce slacktivists or clicktivists who engage with ideas
online but do not engage in offline political actions, such as rallys or
lobbying days. The literature largely criticizes this ``limited''
engagement instead of valuing it as a type of engagement
\citep{Lovejoy2012-pf,Hou2015-gq,Svensson2015-nj,Muralidharan2011-re,Galvez-Rodriguez2014-pz}.
Moreover, collectively this body of scholarship also tends to reinforce certain norms about
the primacy of certain NPO activities over others, such as mobilizing citizens for
rallies, elections or other political actions or fundraising. While these
activities may be primary for established organizations with public
advocacy goals, it is less true of nonprofit organizations as a whole given their wide range of goals and purposes.

In order to investigate why NPOs exhibit the behaviors they do, we studied the
adoption and use of ICTs by 23 nonprofit organizations in Chicago, IL. These organizations vary in size, civic scope, and social reach, and employing an expanded
view of what it means for an organization to ``civically engage''
enabled us to reveal patterns of ICT use that were not found in prior
studies. We use the lenses of infrastructure and moral economy to
explain why the assemblages of tools our participants describe emerged.
We contribute to relevant literature by (a) characterizing these
patterns and (b) articulating how the lenses of infrastructure and moral economies explain the assemblages NPOs create. We conclude by calling for a more situated understanding of NPOs' ICT use in information-related literature, one that embraces the design challenges posed by NPOs' variable needs and contexts, and moves past the notion that there is a ``correct'' or even ``optimal'' way of using ICTs to meet engagement needs. 

\section{Related Work}

\subsection{NPOs and Their Goals}
More than 1.8 million tax-exempt NPOs are registered in the United
States (U.S.) \citep{noauthor_undated-kv} and are
classified by 29 different tax-exempt classifications according to the
Internal Revenue Service (IRS). These nonprofits exist to further a range of
social, cultural, and humanitarian causes 
\citep{noauthor_undated-nn} and advance a variety of goals, including educating local
residents \citep{Kase2008-kj}, improving engagement with stakeholders \citep{Hou2015-gq}, and networking
with the public \citep{Voida2012-fd}. Data from a variety of
sources showcases the scope and scale of these efforts. For instance, the Urban Institute reports that financial contributions to NPOs by individuals, foundations, and corporations totaled \$358.38 billion in 2014. That same year, 25.3\% of adult Americans volunteered personal time, contributing 8.7 billion hours to the needs of shared, community life \citep{McKeever2016-ya}.

Some NPOs use ICTs to support parts or all of their missions.\footnote{We  acknowledge that there is a story to be told about ICT non-use \citep[e.g.,][]{Baumer2014-ec} by NPOs. However, this paper focuses on the technologies NPOs in our study choose to use and how they employed these tools once chosen.} Nah and Saxton \citeyearpar{Nah2012-jc} suggest that these NPO technology choices are motivated by four key
factors: strategy (including fundraising, lobbying and market-based),
capacity (including organizational size, website age and reach),
governance features (including membership, organizations, board size and
efficiency), and external pressures (including donor dependence and
government dependence). Relatedly, Lovejoy and Saxton
\citeyearpar{Lovejoy2012-pf} reveal that NPOs use ICTs for
three primary reasons: information, community, and action. Looking
specifically at Twitter, they find NPOs use it most to engage in
information-related tasks, such as providing updates on organizational
activities to stakeholders. Waters and Jamal
\citeyearpar{Waters2011-bu} also see the result: nonprofits
tend to use Twitter to convey one-way messages rather than exploiting
its dialogic or community building affordances. 

\subsection{NPOs and Technology Choice}
In trying to understand why NPOs choose and use the technologies they do, we turned to literature on technology choice and briefly review major theories below. We focus on three theories that focus on utility and are often cited in information literature.

\subsubsection{Task-Technology Fit}
Much of the technology choice literature in information systems (IS) uses a task framework when explaining how individuals select technologies to use. Tasks are actions that individuals undertake to turn ``inputs into outputs'' \cite[p.~21]{Goodhue1995-im} and are often associated with uses of data collected and provided by a computation system~\citep{Goodhue1998-kx}. Task-technology fit (TTF) is a measure of how well a technology facilitates this input-to-output process for individuals. The TTF model is chiefly concerned with the relationship between technology use and individual worker performance as measured by self-reported indicators of effectiveness, productivity, and performance. When users depend on a system and see a good fit between that system and their tasks, they perceive performance to improve.

Since it was initially proposed as a technology choice framework in 1995, TTF has been applied broadly to investigate a diverse range of information systems and has been combined with or used as an extension of other models related to IS outcomes such as the technology acceptance model (TAM) \citep{Ehrlich2000-pw,Klopping2004-sf}. TTF has also been extended; for instance, \citet{Lu2014-lm} propose a social/task-technology fit (STTF) model, in which the social-technology fit refers to the degree to which a technology (especially social network sites) fits users' social needs. In this model, social characteristics refers to users' needs for social demands. 

\subsubsection{Uses and Gratifications
Theory}\label{uses-and-gratifications-theory}

Instead of basing technology choice on some dimension of task
accomplishment, uses and gratifications theory (U\&G theory) uses a
needs framework to suggest that people actively seek out specific media
to satisfy specific needs. U\&G research has typically focused on how
media are used to satisfy cognitive and affective needs
\citep{Urista2009-mn}. Researchers leverage U\&G theory
to explain what motivates individuals to switch from traditional media
to new media and what kinds of gratifications these media are providing
\citep{Eighmey1998-qp,LaRose2004-vf,Papacharissi2000-uk,Stafford2004-rr}.
A key distinguishing feature of social media are their abilities to
fulfill a need for interactivity \citep{Ha1998-nd}.

\subsubsection{Affordances}\label{affordances}

Researchers argue also that technologies are chosen on the basis of their perceived
affordance(s). The idea of an affordance originates with Gibson, who saw that people relate objects in the world with an
imagined purpose or usefulness \citep{Gibson1984-cb}.
This imagined utility, or the perception of a relationship between an
object and an outcome, is the way that affordances were largely
conceived until Don \citet{Norman1988-sy} moved the idea from the realm of material
objects into the digital world.
\citet{Norman1999-hf} saw that the way an interface was designed would have an
effect on how people thought about its perceived affordance(s). However, Kaptelinin and Nardi \citeyearpar{Kaptelinin2012-tn}
argue that the sense of a technology's potentiality is more than just a
function of the technology itself; it matters in what context or
environment that a tool exists to be used. These authors suggest that the
application of any technology---whether imagined or actualized---must be
understood within a context that gives it meaning. We explain more about
the context in which NPOs choose and employ information and
communication technologies in the section below.

\subsection{NPOs and Engagement in Social Media}
Researchers in a wide range of information-related disciplines have examined NPO social media use, and explicitly address issues of civic engagement via social media. In these studies, ``engagement'' refers to myriad activities including direct advocacy and stakeholder communication. For instance, Hou and
Lampe \citeyearpar{Hou2015-gq} interviewed advocacy
organizations and analyzed their social media feeds with an eye towards
small nonprofits. They argue that social media can facilitate NPO
engagement efforts only if organizations understand their own social
media performance. Similarly, Briones and colleagues' \citeyearpar{Briones2011-uy} study of ICT use
by the American Red Cross 
suggests that a lack of human resources and skills can create barriers
for NPOs trying to use social media to build relationships. Other
studies point out the challenges many nonprofit organizations
face
\citep{Kase2008-kj,Voida2011-tj,Voida2012-fd,Le_Dantec2008-nt}
while recognizing that advanced technology use is often not the highest
priority for small organizations given their other demands (e.g.,
delivering social services)
\citep{Voida2012-fd,Hou2015-gq,Briones2011-uy}. 

Some information-related research criticizes NPOs more directly---usually to say they are not
capitalizing on the interactive or community-building features afforded
by ICTs
\citep{Lovejoy2012-pf,Hou2015-gq,Svensson2015-nj,Muralidharan2011-re,Galvez-Rodriguez2014-pz,Hackler2007-xf}.
Researchers point out that most communication on social
media remains one-way rather than interactive
\citep{Svensson2015-nj} or call for more staff to be
assigned to carry out social media strategies
\citep{Hou2015-gq}. Researchers do recognize that the
constraints NPOs face depend on their membership and resources; specifically
that they rely on volunteers whose expertise may not include
cutting edge ICT use \citep{Le_Dantec2008-nt}. In the most extreme cases, these kinds of criticisms sound eerily like ``blame the user'' arguments
of the past.

Yet, even if we assume that NPOs want to improve their stakeholder
engagement (e.g., to raise funds, to garner political support, etc.),
there is little guidance in the existing research with regard to
measuring social media's influence for this agenda. Most studies focus
on only Facebook and Twitter
\citep{Nah2012-jc,Hou2015-gq}, and they do not present
data on engagement measures beyond dollars raised and signatures
collected. For example, Carboni and Maxwell
\citeyearpar{Carboni2015-nm} sampled five youth development
organizations and found that longer Facebook posts and increased
spending on advertising predict increased stakeholder engagement as
measured by likes, comments, and shares. A higher number of posts
negatively predicts stakeholder engagement, which suggests that frequent
posting is not, on its own, a successful strategy for NPOs to employ.
With the same measurement of engagement, Cho and colleagues
\citeyearpar{Cho2014-ly} explored NPOs' use of Facebook,
finding higher levels of engagement with organizational messages when
two-way symmetrical communication was used, compared to public
information or two-way asymmetrical models. These are measures of input
and interactivity, not impact, and the studies do not address the
motivations and strategy behind visible communications.

Why does this narrow framing of ICT use and civic engagement measurement matter?
The world of nonprofits and their related ICT usage is much broader than
the existing literature would lead us to believe. These studies do not
acknowledge the wide amount of variation in the goals and missions
\citep{noauthor_undated-nn} of the over 1.8 million
nonprofit organizations in the U.S. For instance, health and human
service organizations, religious organizations, and those focused on
hyperlocal geographies (i.e., neighborhoods, small towns) are often
overlooked. These are just a few of the 29 classes the IRS uses to categorize tax-exempt NPOs; the range of activities among organizations without tax-exempt status is also missing. Researchers have already begun to prescribe norms for
nonprofits' technology use and choices. This implies both that NPOs are
making choices from a set of distinguishable, accessible options and
that a single set of norms should apply across NPOs.

Research needs to account for those NPOs that are little more than
volunteer groups with no full-time paid staff---as well as those larger
enterprises, such as the American Red Cross, that employ thousands of staff
members. We also know that not all forms of engagement are the same,
meaning that the way nonprofits choose to use ICTs must also, by
definition, vary accordingly. To these ends, we designed a study to explore this
diverse landscape and to push against what we saw as a premature
institutionalization of nonprofit ICT norms in the information literature. 

\section{Research Study}\label{research-study}

To better understand NPOs' choices about ICTs and their usage, we
conducted an exploratory, qualitative study of 23 different
organizations in Chicago, IL with a specific focus on how these
organizations aim to facilitate civic engagement. This study was
motivated by and must be understood by several methodological guideposts
we oriented our research around. First, we used a broad definition of
civic engagement when selecting nonprofit organizations---namely, we
defined \textit{civic engagement} as, ``the way(s) in which associated groups of
individuals work together to improve the quality of life in their
community(ies).'' Our definition mirrors that of another presented by \citet{Ehrlich2000-pw}. Second, we
intentionally set out to examine nonprofits that were defined by a
variety of organizational missions. While a mission of public advocacy
is a key factor in civic engagement, not all nonprofits consider
themselves advocacy-organizations; to limit our sample to this single
organizational type would be biased. Third, we believe that the
practices of civic engagement extend beyond outreach and
communication-related activities, so we opened ourselves to discovering
more than these standard practices when collecting data. Finally, we chose to include informal and loosely structured groups in our nonprofit
sample to recognize the potentially unique decisions made by volunteers
operating at the grassroots level.

\subsection{Data Collection}\label{data-collection}

To accomplish the goals of this study, we employed a stakeholder-focused
approach. Employing this approach means that we did not assume Chicago
nonprofits possessed the structures that are associated with
traditional forms of organization. Instead, we recruited participants
from organizations based on the concept of community attachment. By
\textit{community attachment} we mean the interpersonal, participatory, and
sentimental connections people have to their communities
\citep{Kasarda1974-us}. This provided a novel way of
examining nonprofits rooted in civic engagement literature, which argues
that community attachment and social capital are two mechanisms that
foster civic engagement
\citep{Fieldhouse2010-pi,Halpern2004-bj}---that more
personal connections, participation, and positive sentiment about one's
community encourage communal activities.

We recruited interview participants through email and Twitter in the
fall of 2017. Potential participants came from a directory of
nonprofits. We enrolled a total of 23 individuals who were employed by,
or affiliated with, 23 Chicago nonprofit organizations, which we detail
further below. Fourteen participants were men and 9 were women; 18 were
White, 3 American-Indian, 1 Black, and 1 Asian. The majority of study
participants belonged to more than one nonprofit or volunteer
organization, so we asked them about the organization to which they felt
most attached. Subjects reported that they were most attached to
nonprofits where they spent the greatest amount of time, which
indicates our data was likely to be accurate. We do not include a table of
participants here but instead provide \textit {Table~\ref{tab:org-types}} that summarizes the kinds
of organizations with which they were affiliated. Because we used IRS data to recruit participants, we believe most of the organizations were NPOs in the tax-exempt sense. Though, we did not confirm the tax status of any of the organizations our participants discussed, the NPOs in our study serve their ``quality of life'' missions through direct advocacy, religious practice, social events, and other activities.

Our interview subjects represented organizations that possess a wide
variety of missions, ranging from grassroots political organizing to
promoting literacy through reading. To better understand the overall
sample, we identified six broad types of organizations according to
their primary missions, which we also define and enumerate in \textit {Table~\ref{tab:org-types}}. Our
sample was neither random nor representative but instead included a wide
variety of NPOs to support claims about which patterns likely hold and
those which do not in terms of their missions. One commonality shared across the whole sample was each organization sought to improve the 
``quality of life'' of the communities they served. We conducted semi-structured interviews with each of our 23 NPO
participants. During these interviews we asked questions related to the
concept of community, offline communities, tools used by nonprofits to
communicate with the public, civic engagement, community attachment,
information access, tool adoption, and interaction with others online.
Each interview was conducted by a graduate student in a face-to-face
setting, was audio recorded, and then transcribed by a third-party
service.

\begin{table}
\caption{Categories of the 23 NPOs with whom our participants
were affiliated}
\label{tab:org-types}
%\begin{minipage}{\columnwidth}
\begin{center}
\begin{tabular}{>{\raggedright\arraybackslash} p{6em} p{24em} p{6em}}
\toprule
Type & Description & NPO Count\tabularnewline
\midrule
Advocacy organization & Raised awareness about issues among stakeholders
and pushed for changes like criminal justice reform & 3\tabularnewline\hline
Social group & Dedicated to creating or maintaining social connections &
2\tabularnewline\hline
Interest group & Non-political groups motivated by shared interests &
5\tabularnewline\hline
Political community & Sought to create electoral coalitions of
individuals to support candidates in efforts to win office and pass laws
& 4\tabularnewline\hline
Religious community & Motivated by religious affiliations &
1\tabularnewline\hline
Residential community & Dedicated to issues within residential
geographies such as traffic congestion and gentrification &
8\tabularnewline
\bottomrule
\end{tabular}
\end{center}
\bigskip\centering
%\end{minipage}
\end{table}

\subsection{Data Analysis}\label{data-analysis}

After conducting our interviews, we coded them in Dedoose using both an
inductive structural \citep{Saldana2015-fs} and
literature-driven approach to develop codebooks related to ICTs, affordances, and adoption causes.
Our ICT-related codebook was created by listing all ICTs (N = 56)
mentioned in interviews. We used literature\footnote{See, for instance,
  \citep{Auger2013-xb,Davis1989-cj,Guo2013-eo,Nah2012-jc,Rogers2003-bd,Smith2015-cz,Walker1969-wn,Park2009-qu,Goodhue1998-kx,Goodhue1995-im,Hou2015-gq,Kase2008-kj,Voida2012-fd,Lovejoy2012-pf} and the technology choice section above.}
to guide the development of our affordance and adoption-related
codebooks. In particular, this meant we itemized discrete concepts from
theories of technology choice and combined them with the findings of
prior NPO adoption studies. To determine the codes included in our
codebooks, we combined duplicate concepts based on a process of reaching
shared consensus. In the end, we labeled 20 affordances and 15 adoption
causes. We provide the complete list of codes and examples of passages coded according to these three codebooks in \textit{Appendix A}.

We applied codes to full conceptual units at the sub-paragraph level. We
did not limit the number of codes per passage, but we required concepts
to be explicitly stated or strongly implied in the text. For example,
speaking about why he used an email list to work with a political group,
one interviewee said that, ``Whoever makes the list will automatically
put all our membership on the list and it blasts out to everybody''
(P19). In this statement, the interviewee indicated that group members
used a tool because other people signed them up for it. This passage was
coded as an adoption-cause related to \textit{Leadership}.

Finally, to provide an additional level of granularity to study
findings, we exported coded data from Dedoose and tabulated
co-occurrence counts. Examining code co-occurrences provided a way to
examine relationships between codes in passages of text, such as
Facebook's use as a tool to share links to news articles. To account for
differences across participants in the number of tools they mentioned,
and how many times they mentioned them, we normalized all co-occurrence
counts. This produced a score for co-occurrence groupings that ranged
between ``0'' and ``1'' and reflected the proportion of total
occurrences relative to the larger code category. We calculated
co-occurrences in three tables: ICT x Affordances and Adoption Causes;
Affordances and Adoption Causes x Affordances and Adoption Causes; and
Affordances x ICTs. A natural break occurred around 15\% (or 0.15), so
we discounted co-occurrences below that threshold to make data analysis
more meaningful. We chose that threshold because it lessened the
potential for rare code co-occurrences to appear more meaningful than
they actually were. Next, we analyzed the co-occurrence of codes from
all three codebooks, which provided a measure of how frequently,
relative to all affordance and adoption references, a particular ICT was
discussed.

\section{Findings }\label{findings}

The analysis of our interviews revealed a complex set of relationships
among ICTs, the affordances our participants understood them to have,
and articulated rationales for their adoption. In this section, we
discuss ICTs, affordances, and adoption causes frequently mentioned in
our interview transcripts. Code applications are shown in \textit{Table~\ref{tab:code-freq}}, which
makes clear that Facebook dominated the other ICTs that nonprofits and
their affiliates reported using. With regard to affordances, five
perceived uses accounted for most (61.4\%) interview passages that we
coded in our data. Finally, NPOs articulated five key explanations for
their ICT adoption choices. Within our sample, these rationales
accounted for 72.5\% of relevant, coded interview passages. At the end
of the section, we describe how NPOs assembled technologies for situated
needs and report on value-based technology choices.

\begin{table}
\caption{Frequencies with which codes were applied}
\label{tab:code-freq}
%\begin{minipage}{.75\textwidth}
\begin{center}
\begin{tabular}{>{\raggedright\arraybackslash} p{6em} p{24em} p{6em}}
\toprule
Codebook & Code & Applications \tabularnewline\hline
\midrule
ICT & Facebook & 702 \tabularnewline\hline
ICT & Twitter & 266 \tabularnewline\hline
ICT & Email & 192 \tabularnewline\hline
ICT & Facebook Events & 128 \tabularnewline\hline
ICT & Websites & 106 \tabularnewline\hline
Affordances & Sharing links, media, and other information &
343 \tabularnewline\hline
Affordances & Advertising and promoting information deemed valuable by
nonprofits & 298 \tabularnewline\hline
Affordances & Finding and retrieving information & 280 \tabularnewline\hline
Affordances & Organizing and coordinating events & 269 \tabularnewline\hline
Affordances & Fostering a sense of presence or attachment &
146 \tabularnewline\hline
Adoption causes & Perceived benefits & 362 \tabularnewline\hline
Adoption causes & Cultural and personal attitudes & 196 \tabularnewline\hline
Adoption causes & Nonprofit goals and strategies & 129 \tabularnewline\hline
Adoption causes & Audience composition & 120 \tabularnewline\hline
Adoption causes & Perceptions about ICT ease of use & 105 \tabularnewline
\bottomrule
\end{tabular}
\end{center}
\bigskip\centering
%\end{minipage}
\end{table}

\subsection{The Predominance of
Facebook}\label{the-predominance-of-facebook}

All of our participants echoed the sentiment expressed by one
interviewee with regard to Facebook: ``{[}It{]} is definitely number one
just because it's sort of like the default. It's like the standard, you
know social media that everything else is, sort of like, measured by''
(P18). Participants also frequently talked about Facebook walls and/or
pages, using Facebook to organize events, communicating (in private)
with individuals through Messenger, and creating groups to coordinate
activity. Looking at our co-occurrence tables, the code \textit {Facebook} co-occurred with all 20 possible affordances and 14 of 15 possible ICT
adoption codes. Unsurprisingly, the Facebook event code co-occurred with
organizing and coordinating events. Interviewees also said that they
used Facebook's event functionality because they were already familiar
with the tool. Even the integration of technologies in assemblages,
which we discuss below, was seen as a Facebook-related boon. P23
elaborates:

\begin{quote}
If someone was posting a photo on Instagram, that photo would show up on
Facebook and it would show up on our website and in our feed, or something
like that, and it would also mention and promote other organizations
that we're collaborating with more often. So {[}by doing that\ldots{} we
can ideally{]} take advantage of the publicity that another organization
might do as a result, and that could increase the number of people who
would see it.
\end{quote}

The reasons participants reported using Facebook were multifaceted, 
but they often related to audience reach. Discussing ICTs in relation to political
recruitment and organizing, P6 said: `'Facebook makes it easy for people to
invite their friends {[}to our events, because people already\ldots{]} spend 
a lot of time on Facebook.'' Participants believed that ``{[}almost{]} everyone
is on Facebook'' (P16) and said they adopted Facebook because it provides 
access to ``a wider audience'' (P1) than competitors, regardless of their engagement 
needs. One participant went so far as to call Facebook ``the universe'' 
(P5), referring to the many functions that it afforded. 
Yet, ``reach'' was not the only thing NPOs cared about when communicating with members of the public. Elaborating on this point, P20 said:
 
\begin{quote}
We're looking to retain the attention of people who already support our issue
but also making things easy enough to understand that it's accessible to a larger 
audience. While we don't compromise our views to reach a wider audience, we
do try to use that space to really, really amplify our messaging in a way that is 
accessible to people who are already plugged in.
\end{quote}

Finally, while Facebook emerged as a central topic in our interviews, it was clear to us
that it could not meet NPO needs in all circumstances.

\subsection{Assembling Alternatives for Situated
Needs}\label{assembling-alternatives-for-situated-needs}

The other ICT applications mentioned by interviewees were often employed
by NPOs to improve on one of Facebook's identified weaknesses. For
example, in interviews, a near universal complaint about Facebook was
that its RSVP function did not accurately predict how many attendees
events would have. NPOs ``want to know people are actually going to be
there and not just clicking like'' (P20) on events. In response,
alternative tools like Eventbrite and Evite were used to achieve more
accurate head counts. Participants suggested these tools were more
accurate because using them to RSVP required marginally more investment
from the audience to complete the RSVP form or book a ticket (even if
free) than clicking ``yes'' or ``interested'' on Facebook:
``{[}EventBrite is{]} much more tangible and much more of a commitment
than just clicking a button on Facebook'' (P04).

A ``long tail'' of 46 different tools such as phones/SMS texts,
Instagram, Facebook Groups, and Facebook Messenger, EveryBlock, Slack,
etc. were also mentioned by participants as alternatives or extensions
of Facebook, but three in particular---Twitter, Email, and
Websites---were commented on the most for their particular affordances.
Twitter was articulated as a popular platform for sharing ``geopolitical
stuff'' (P18) and ``one-liners'' (P19). Twitter's hashtags were also
seen to have a particular utility, as P13 commented: ``you can look up
the hashtag. You can search the hashtag or follow it. There are some
people there at the event also posting at the same time.'' Participants
noted email's value as a reliable way to contact individuals within the
organization: ``The email is for my boss, the email is for volunteers,
the email is not for people in general to the community'' (P02). Finally,
websites were usually spoken about as specific sources of information
that offered NPOs more control over their virtual presence. One
participant juxtaposed her employer's website with Facebook by saying
that, ``we're able to put more detail on our website. {[}It allows 
us\ldots{}{]} to control who sees what and when'' (P23).

The articulation of technology affordances also occurred in relation
to ICTs in combination. Perceived affordances emerged from tool
assemblages that were created by our participants to accomplish specific
goals, sometimes used in specific sequential patterns. Discussing this,
one interviewee talked about the final stage of a five-tool process used
to coordinate events:

\begin{quote}
I would say that our email blast is our final funnel. We get people who
learn about an event on Facebook and come to the event, but we are
casting a wide net. Once we get your email we know that you're actually
interested. Then we can communicate very directly about the stuff we're
doing and the priorities we have going. (P11)
\end{quote}

NPOs appeared to be assembling ICTs together to create a viable
means---according to them---to advertise, coordinate, and organize
events. NPOs also expressed an interest in employing this strategy more
generally. Talking about leveraging multiple tools to meet
organizational goals, one participant said, ``I think that if we could
do anything we would plan in advance our strategy and think about
{[}this type of assemblage\ldots{]} more purposefully'' (P23).

\subsection{Value-Based Technology
Choices}\label{value-based-technology-choices}

Throughout our interviews we also saw evidence that individuals'
personal attitudes affected their thinking about technologies and
decisions whether or not to use them to advance goals. For example,
speaking about Twitter, one interview subject felt that it was
``boring'' (P5). Similarly, Snapchat was considered a tool used by a ``younger audience {[}than ours{]}'' (P21). But a much more
predominant insight found in our data was that organizations signaled
their values when making their ICT choices. One example of this is the
way that NPOs went out of their way to ensure communications remained
private. Participants in our sample who worked with undocumented
immigrant communities and environmental activists that protested the
Dakota Access Pipeline described choosing to use WhatsApp because of its
encryption capability. Another participant spoke about community organizing and inclusivity as a motive for ICT adoption: ``I think Facebook is the easiest way for
people to organize themselves but there's also a barrier with who can and 
cannot {[}\ldots get involved{]}'' (P8). Elaborating further, this ``barrier'' was revealed to be an inability by non-English speakers to read a neighborhood association newsletter. In response, the participant's brother created a Spanish Facebook page for the association to help promote neighborly inclusivity. 

Even technology non-use was articulated through a value-based lens. In speaking 
about a progressive political group that used NationBuilder to register and
organize voters, P6 remarked that the company that sold it ``took a bunch of
credit for Trump winning {[}the 2016 election{]}.'' In response, the NPO planned to stop using
NationBuilder once their annual subscription ended; they did not want to  patronize a company that served a key political antagonist. Finally, in discussing why she doesn't
use Facebook or email, one participant (P20) commented:

\begin{quote}
{[}A{]} lot of the people that we're trying to help\ldots{} they're coming out of jail and they don't have cell phones. If they do have a cell
phone, it's a government phone and they're not able to access anything.
Or they don't have a computer\ldots{} We actually sent snail mail out to
the people that we bonded out.
\end{quote}

In short, NPOs chose ICTs based on how they aligned or did not align
with certain social, political, and cultural values the organizations and their members held. It also bears mentioning that these values align with civic missions typical of the nonprofit sector.

Concluding this section, our participants assembled combinations of ICTs
that can be described generally as ``Facebook+''. In these
assemblies, the Facebook platform intersected with all (or nearly all)
affordances and adoption causes identified by prior literature and
mentioned by our participants. The choices NPOs made to extend and
augment Facebook aligned with their values and their stakeholders' needs. The long tail of ``+'' ICTs included tools such as EventBrite and Evite that serve particular RSVP purposes, WhatsApp and cell phones that offer less surveillance (and require that stakeholders possess less
sophisticated devices), and that enabled NPOs to control their message
and public image. In the next section, we explain how these assemblages emerged and why prior discussions of technology choice among NPOs miss
important infrastructural and moral considerations that NPOs take into
account.

\section{Discussion}\label{discussion}

NPOs in our study relied on Facebook as infrastructure for their communication and outreach, and they chose ICTs that fit their values  and resource constraints. Broadly speaking, existing theories of technology choice emphasize a utility model, whether that utility is expressed in terms of affordances \citep{Norman1999-hf,Kaptelinin2012-tn}, uses and gratifications \citep{Papacharissi2000-uk,Urista2009-mn}, or a perceived synergy of some kind. These theories do
not adequately explain the assemblages that NPOs describe using in our data.
Instead, we argue that NPO ICT use is better explained using the dual lenses
of infrastructure and moral economy. While we are not suggesting that NPOs pay no need to technological efficacy, we do want to raise the notion that their choices appear to be equally motivated by community practices, standards, and expectations. Existing theories of technology choice suggest the technology and what it can do is the most prevailing concern when users weigh their choices. By contrast, we found that how the technology is embedded in the NPOs' worlds and how it aligns with their values were the most salient factors in guiding their choices.

\subsection{Facebook as NPO
Infrastructure}\label{facebook-as-an-npo-infrastructure}

The traditional view of infrastructure articulates a sociotechnical substrate on which other tools and systems are built, used and maintained according to community standards and practices. As such, infrastructure is a fundamentally relational entity
\citep{Jewett1991-nd} that emerges (and perpetually re-emerges) in practice
\citep{Star1996-nh}---a fact that many researchers have already pointed out
\citep{Erickson2016-vd,Ribes2010-ty,Ribes2014-tj,Star1996-nh}.
For the NPOs in our study, Facebook possesses all of these
characteristics: it is a sociotechnical, relational substrate upon which NPOs create
bespoke assemblages. This situated and ongoing practice of Facebook use among NPOs aids in evolving and expanding infrastructure over time while at the same time allowing NPOs to meet their engagement goals. Other studies of NPO technology use also found assemblages rather than single or even primary tools \citep{Voida2011-tj,Stephens2007-fn}, but our participants describe Facebook as the infrastructure on which these assemblages are built and not just another tool in the set.

Several empirical insights from our data underscore the framing of
Facebook as infrastructure. First, it is embedded in the social
arrangements of nonprofit organizations, is used frequently, and supports
the ICT needs of nearly every required NPO task. Speaking about this point,
one participant said, ``I think that {[}Facebook{]} is pretty darn
complete. I mean, they got the Messenger. You can direct message people.
You can invite people. And you can just post publicly'' (P3). The centrality of Facebook within the NPOs' communities further underscores its infrastructural nature. There is something imperative about it, which led one group to force a member to ``make a Facebook {[}account{]}'' even though she did not want one (P14). However, usage here is not merely a matter of capitualization to social pressure. Civic engagement requires interacting with community members, and the most straightforward way to do so, according to our interview participants, was to use the same tools (e.g., Facebook) that their stakeholders use. \citet{Nemer2017-gq} found that people became more active citizens when they were comfortable using technologies. NPOs appear to recognize that their stakeholders use Facebook, are comfortable there, and may intuitively leverage that confidence to increase civic engagement. The larger social practice of using Facebook provided NPOs audience reach that they would not otherwise have. Coupled with the functions associated with Facebook as a platform (e.g., providing event details, sharing information), this reinforces its centrality as infrastructure embedded in NPO social arrangements.

Complementing Facebook's practice-oriented ubiquity in NPOs' toolkits, however,
is an even more important point. If not always technically true (i.e.,
not necessarily via API), Facebook acts as an installed base upon which
NPOs add on related technologies to fill gaps or extend desired
features. For example, only when Facebook failed to provide accurate
RSVP counts did one NPO turn to RSVP tools such as EventBrite and Evite
to fill their needs. Likewise, ICTs such as Instagram or NationBuilder
were also chosen to augment Facebook with particular functionalities,
not to replace it. Twitter was agreed to be ideal for ``spur of the moment observations'' (P15), which meant that it was used to live-tweet events usually publicized on Facebook. Recalling Star and Ruhleder's \citeyearpar{Star1996-nh} 
dimensions of infrastructure, we claim that
Facebook is NPO infrastructure because it plugs in to
other ICTs (or allows them to plug into it). Importantly, this extensibility, as shown by our examples, reinforces the embedding of Facebook into NPOs' practices rather than disembedding it into a series of separate technological moves. Seen together, these  infrastructural maneuvers provide NPOs with tremendous impact beyond what a single tool could provide.

This finding is in line with a study of volunteers and their technology
use by \citet{Voida2015-db}. They find, similarly, that volunteers in
nonprofits employ technologies (e.g., productivity software, vehicles)
that are ``infrastructural already'' (p. 12). In other words, volunteers seek
out and use everyday tools that are extensible enough to accommodate
their needs, not tools that are specifically designed for nonprofits.
Our participants experienced Facebook in the same way---they leveraged it
as a multipurpose infrastructure for communication and interaction, and
extend and augment it with other ICTs as needed. Voida and colleagues ask us to imagine technologies that ``include dimensions of work and social structure'' (p. 12), and our participants describe Facebook as a boundary-crossing, transecting infrastructure. Existing theories of
technology choice do not adequately explain how and why NPOs make these
`Facebook+' decisions. Research on motivation in social media use suggests that motivation varies among both social media tools (e.g., Facebook, Twitter) and social context~\cite{Oh2015-qe}; our participants talked about their motivations in explicitly principled language. Thus, we argue that the lens of moral economies,
described next, explains these practices with greater parsimony than
existing research.

\subsection{Technology Choices in Relation to Moral Economies
}\label{technology-choices-in-relation-to-moral-economies}

Our second finding reveals that technology choice by
the nonprofits in our sample was driven largely by a sense of moral fit and resource accessibility.
These ideas derive from a moral-economic perspective, which posits that economies are organized systems of resource exchange with psychological and normative regularities
\citep{Daston1995-ce}. Scott develops the notion of a
moral economy using the example of peasants. When viewed in moral economic terms, peasants exist within a system that has a well-defined ``notion of economic justice'' and a 
``working definition of exploitation''
\citep{Scott1977-yn}. As such, it should come as little surprise if they do not rebel under seemingly obvious exploitative conditions in search of income maximization---the rationalized economic expectation---but instead express a logical agency by focusing on creative forms of subsistence. Drawing on this perspective, Vertesi and
colleagues \citeyearpar{Vertesi2016-gp} apply a moral economy lens to
understand how people make decisions in personal data management. They define ``the moral economy
of data management'' as ``a locally adjudicated way of combining
devices, services, and social ties so as to personally embody a good and
appropriate relationship to personal data'' (p. 479). In both of these cases, the moral economic move by actors stands in contrast to the purely economic (i.e., utility-driven) move. It is this local adjudication, which balances resource choices with local values, that explains the actors' context-specific system choices. 
 
In the moral economy in which the NPOs in our study operate, we see the same normative decision frameworks in play as organizations seek to align their technology choices with their values and maximize localized resources at hand (i.e., social capital, legitimacy). In so doing, we see them choosing to exploit
existing systems or use affordable ICTs that do not require dedicated or
specialized technical resources as a creative and intentional means to connect with constituents or affect certain desired social outcomes. Similar to the individuals in Vertesi and colleagues study, the NPOs in our study use technology to enact a complex vision of relationality that is normatively appropriate and sufficiently impactful. For example, NPOs consciously aim to reach and empower their constituents, but do so with regard to minimizing surveillance (e.g., NPOs supporting undocumented immigrants by using WhatsApp) or matching communication preferences (e.g., avoiding ShapChat or utilizing postal mail). In other words, these NPOs are crafting infrastructural assemblages that both respect their constituents' privacy needs and social media routines while also allowing them to meet their responsibilities to their stakeholders.

The NPOs we talked to are also similar to the peasants in Scott's~\citeauthor{Scott1977-yn} discussion---they are struggling to survive first. Participants explicitly mentioned day-to-day  activities (e.g., announcing events, sharing news stories) or modified their comments with phrases such as "just trying to share" (P04) or "just trying to get people in [the space]" (P01) in ways that illustrate their attempts to meet their basic needs. They also talk about leveraging social media's reach to facilitate the creation of offline relationships, but did not mention social media as the end goal or final site of engagement.

In sum, the NPOs in our study demonstrated complex moral intentionality in their ICT choices. They were not driven by utilitarian motives to maximize economic activity or donations---participants rarely talked about fundraising. They used Facebook not out of isomorphic social pressure, or as underresourced agents, but as creative actors who saw a way to exploit a ubiquitous tool by refashioning into an infrastructural assemblage with bespoke ``gap fillers'' such as EventBrite or WhatsApp. These choices recognize that certain social media enable them to access and then maximize the attentional and social resources of their community. These findings are in line with earlier research about civic engagement and ICTs in Chicago that found distinct communities selected different technologies for discussing crime in part because of their various levels of trust in the police and fear of retaliation \citep{Erete2014-xj}. Erete and colleagues \citeyearpar{Erete2016-su} also found that Chicagoans adapted different technologies for reaching different audiences (e.g., using email to communicate with police) or holding public officials accountable (e.g., recording meeting notes to capture public officials' verbal statements). In their study and ours, Chicagoans used technologies that supported their values---privacy, accountability---not just revenue or engagement. 

As we noted earlier, some of the research on
NPO ICT use takes a ``they should do X'' tone in discussing how NPOs
use ICTs (e.g., ``Small  organizations  need  to  better 
understand and evaluate the  success of their  social  media  performance'' \citealp{Hou2015-gq}). These authors assume that NPOs are trying to maximize engagement online and that engagement is a primary goal. But
what if NPOs, like peasants, are trying to subsist? What if their
primary concern is not engagement but survival because survival is
necessary for their other goals? Sensitivity to these concerns is paramount for researchers, technology
designers, and nonprofits. As Le Dantec and Edwards
\citeyearpar{Le_Dantec2008-nt} warned us 10 years ago, we
must be careful to be supportive rather than disruptive when encouraging
ICT use in nonprofits. NPOs operate in conditions of resource
constraint, sometimes including minimal technical expertise, with
majority-volunteer workforces who often attempt to serve already
marginalized populations. In our research, we witnessed NPOs appropriating existing
infrastructure and extending it in line with their values---a practice that reveals commendable adaptability. Though we advocate for a empathetic reading of their activities, we recognize that researchers can also encourage NPOs to think about how social media may advance or change their activities in ways they haven't considered.

\section{Conclusion}\label{conclusion}

We set out to understand how NPOs make choices about technologies to use
in civic engagement activities and found that all their choices now flow
through or at least contend with Facebook. The accounts our participants
provide reveal that NPOs leverage ICTs within their 
local contexts, the financial and expertise
constraints they face, and the information infrastructure that Facebook
has become. Because existing utility-based theories of technology choice do not
adequately explain the behaviors we see, we use the lenses of
infrastructure and moral economies to explain the emergence of
assemblages of tools our participants articulated. In doing so, we
highlight Facebook's embeddedness in the NPO universe and clarify why
particular patterns of tools and uses appear. We argue that these
particular assemblages are a product of Facebook's infrastructural
position in contemporary communication systems and NPOs' values generally conceived, and
not, as prior work suggests, a failure of NPOs to appropriately
capitalize on technical features of ICTs in a functional sense.

\section{Acknowledgements}
This work was supported in part by the National Science Foundation under Grant No. 1822228. We are grateful to Xi Rao and Anders Finholt for their help conducting and coding interviews.

\appendix

\section{Additional Details about Coding Process and Codebook}
\begin{spacing}{1.0}
\begin{longtable}{| p{.20\textwidth} | p{.20\textwidth} | p{.20\textwidth} | p{.40\textwidth} |} 
%\begin{minipage}[t]{\textwidth}
%\begin{center}
%\begin{tabular}{>{\raggedright\arraybackslash} p{4em} p{4em} p{4em} p{24em}}
\toprule
ICT & Affordance & Adoption & Excerpt\tabularnewline
\midrule
Facebook & Foster Sense of Presence or Attachment & Perceived Ease of
Use & "Facebook has made it so much easier to form groups and to meet up
and to have conversations with multiple people. I feel like I wouldn't
even have met a lot of these people in groups if it hadn't been for
Facebook."\tabularnewline\hline
Facebook Event & Organize and Coordinate Events & Leadership & "I don't
make that decision. The organizer will make that decision. I wouldn't
mind if they created a Facebook event for that, but I guess it's just
because it's a professional type of meeting is how I would describe
it."\tabularnewline\hline
Facebook Messenger & Personal Use & Audience Composition & "I just use
Facebook Messenger for two or three friends that I talk to that are
really close friends that we just talk everyday kind of a
thing."\tabularnewline\hline
Facebook & Political Discussion and Organizing & --- & "If I support
Planned Parenthood, for example, and I'll follow them on Facebook and if
they say on their local Chicago page, `Hey, there's a march in downtown
Chicago,' then I'll participate in that."\tabularnewline\hline
Facebook Group or Poll & Present Opinions & Goals and Strategies & "For
the book club, we use Facebook to discuss what book we want to read.
Sometimes we'll run a poll. {[}...{]} like, `Hey here are some
suggestions, which book do we want to read?'"\tabularnewline\hline
Facebook & --- & Audience Size & "I guess just there's more people on
there, more people communicating, it's more active. So that makes me
want to use it more."\tabularnewline\hline
Email & Distributed Work & --- & "Agendas are usually sent out the day
before via email. We do make use of email lists too."\tabularnewline\hline
Instagram & Encourage Stakeholder Interaction & --- & "So, if someone
was posting a photo on Instagram, that photo would show up on Facebook.
It would show up {[}...{]} in a feed or something like that, and it
would also mention and promote other organizations that we're
collaborating with."\tabularnewline\hline
Google & Find and Retrieve Information & --- & "Yeah, it depends on how
I feel, I don't know. I don't know how I make that decision. Like, I
want news right now about something, I would Google it."\tabularnewline\hline
Snapchat & Personal Use & Relationship With Other Tool Users & "I don't
have that many friends on Snapchat. So, it's not very useful for
me."\tabularnewline\hline
Twitter & Share Links, Media, and Other Information & Cultural and
Personal Attitudes & "{[}Twitter isn't{]} useless, I just don't
understand it. My brother uses it a lot for like different articles. He
uses it to share different articles and opinions."\tabularnewline\hline
Phone/SMS Text & --- & Audience Composition & "It's just texting or
calling {[}to communicate and organize{]} because they're older, so
they're not on social media."\tabularnewline\hline
Website & --- & Resources & "Part of this is like, again, having a web
developer that is a volunteer. We can't be like, 'We need you to make
the thing and do it now because we paid you.' No. We love and respect
this person and get that they're overwhelmed."\tabularnewline\hline
NationBuilder & --- & Values and Ethical Considerations & "A bunch of
the people {[}\ldots{}{]} are really upset about NationBuilder, because
they took a bunch of credit for Trump winning. They sold their product
to various Trump-aligned interests."\tabularnewline\hline
--- & Advertising and Promotion & Perceived Benefits & "Mainly just
marketing I guess. Just trying to get people to get in the coffee shop.
You know if I offer a special or {[}...{]}. Yeah, just trying to get
people in the door, advertising."\tabularnewline
\bottomrule
%\end{tabular}
%\end{center}
%\bigskip\centering
%\footnotesize\emph{*}Included for descriptive purposes. Not among the top-10 most
%frequent ICT codes that were applied to transcripts.
%\end{minipage}
\caption{Representative interview quotes with code applications}
\label{tab:codebook}
\end{longtable}
\newpage

\begin{longtable}{| p{.33\textwidth} | p{.33\textwidth} | p{.33\textwidth} |} 
%\begin{minipage}[t]{\textwidth}
%\begin{center}
%\begin{tabular}{>{\raggedright\arraybackslash} p{4em} p{4em} p{4em} p{24em}}
\toprule
ICT & Affordance & Adoption \tabularnewline
\midrule
Action Network & Audience Composition & Advertise or Promote \tabularnewline\hline
Bandcamp &	Audience Size & Archive \tabularnewline\hline
Blog & Cultural and Personal Attitudes & Build Organizational Capacity \tabularnewline\hline
Camera/Videocamera & Dependencies & Collect Data \tabularnewline\hline
CiviCRM	& Familiarity with Tool & Creativity and Experimentation \tabularnewline\hline
Constant Contact & Goals and Strategic Orientation & Distributed Work \tabularnewline\hline
Craig's List & Leadership & Encourage Stakeholder Interaction \tabularnewline\hline
Doodle & Other & Find and Retrive Information \tabularnewline\hline
Email & Perceived Benefits & Foster Sense of Presence or Attachment \tabularnewline\hline
EventBook & Perceived Ease of Use & Lobby Officials \tabularnewline\hline
Eventbrite & Relationship with Tool Users & Motivate Stakeholders \tabularnewline\hline
EveryBlock & Resources & Network \tabularnewline\hline
Evite & Stakeholder Influence & Organize and Coordinate Events \tabularnewline\hline
Facebook & Urgency & Personal \tabularnewline\hline
Facebook Event & Values and Ethical Considerations & Political Discussion and Organizing \tabularnewline\hline
Facebook Group or Poll &  --- & Present Opinions \tabularnewline\hline
Facebook Messenger &  --- &	Privacy \tabularnewline\hline
Flickr &  --- & Real-time Event Discussion \tabularnewline\hline
Flyers & --- & Share Links, Media, and Other Information \tabularnewline\hline
GoFundMe & --- & --- \tabularnewline\hline
Google & --- & --- \tabularnewline\hline
Google Drive/Docs & --- & --- \tabularnewline\hline
Google Other & --- & --- \tabularnewline\hline
GroupMe & --- & --- \tabularnewline\hline
Hype Machine & --- & --- \tabularnewline\hline
Instagram & --- & --- \tabularnewline\hline
LinkedIn & --- & --- \tabularnewline\hline
MailChimp & --- & --- \tabularnewline\hline
MeetUp & --- & --- \tabularnewline\hline
MeisterTask & --- & --- \tabularnewline\hline
MySpace & --- & --- \tabularnewline\hline
Newsletter/Newspaper & --- & --- \tabularnewline\hline
Nextdoor & --- & --- \tabularnewline\hline
OkCupid & --- & --- \tabularnewline\hline
Phone/SMS Text & --- & --- \tabularnewline\hline
Pintrest & --- & --- \tabularnewline\hline
Postal Mail & --- & --- \tabularnewline\hline
Reddit & --- & --- \tabularnewline\hline
Server & --- & --- \tabularnewline\hline
Skype/Videoconferencing & --- & --- \tabularnewline\hline
Slack & --- & --- \tabularnewline\hline
Snapchat & --- & --- \tabularnewline\hline
SurveyMonkey & --- & --- \tabularnewline\hline
Television/Radio & --- & --- \tabularnewline\hline
Tinder & --- & --- \tabularnewline\hline
Tumbler & --- & --- \tabularnewline\hline
Twitter & --- & --- \tabularnewline\hline
Unspecified & --- & --- \tabularnewline\hline
Viber & --- & --- \tabularnewline\hline
Website & --- & --- \tabularnewline\hline
WhatsApp & --- & --- \tabularnewline\hline
Yelp & --- & --- \tabularnewline\hline
YouTube & --- & --- \tabularnewline
\bottomrule
%\end{tabular}
%\end{center}
%\bigskip\centering
%\footnotesize\emph{*}Included for descriptive purposes. Not among the top-10 most
%frequent ICT codes that were applied to transcripts.
%\end{minipage}
\caption{ICT, affordance, and adoption-related codebooks, in alphabetical order}
\label{tab:codebook}
\end{longtable}

\end{spacing}

% Bibliography
%\bibliographystyle{apacite}
\bibliography{npo-refs}

\end{document}